\documentclass[letterpaper]{article}

\usepackage{aaai}
\usepackage{amsmath}
\usepackage{amssymb}
\usepackage{color}
\usepackage{tikz}
\usepackage{hyperref}

\newtheorem{theorem}{Theorem}[section] 
\newtheorem{lemma}[theorem]{Lemma} 
 
\newtheorem{proposition}[theorem]{Proposition} 
\newtheorem{definition}[theorem]{Definition} 
\newtheorem{example}[theorem]{Example} 

\definecolor{shadecolor}{rgb}{.9,.9,1}
\newcommand{\defining}[1]{\textbf{#1}}
\newcommand{\bool}{\mathit{Bool}}
\newcommand{\false}{\mathit{false}}
\newcommand{\true}{\mathit{true}}
\newcommand{\nat}{\mathbb{N}}

\newcommand{\integ}{\mathbb{Z}}

\newcommand{\zerotol}{{[0..l]}}

\newcommand{\coeffSum}[2]{\sum_{i=#1}^{#2}a_ix_i}
\newcommand{\coeffSumPrime}[2]{\sum_{i=#1}^{#2}a'_ix_i}

\newcommand{\OP}{\mathit{OP}}
\newcommand{\Cut}[3]{\mathit{S}(#1,#2,#3)}
\newcommand{\CutT}[2]{\mathit{S}(\cdot,#1,#2)}
\newcommand{\mint}{s}
\newcommand{\maxt}{b}
\newcommand{\degree}[2]{(#1,#2]}
\newcommand{\LPBT}[2]{}

\newcommand{\sd}[1]{\colorbox{shadecolor}{#1}}
\newcommand{\ld}[1]{\raisebox{-1ex}[0cm][0cm]{\sd{#1}}}
\newcommand{\ud}[1]{\raisebox{1ex}[0cm][0cm]{\sd{#1}}}
\newcommand{\ub}[1]{\raisebox{1ex}[0cm][0cm]{#1}}

\nocopyright
\begin{document}

 \title{Implementations of two Algorithms for the\\ 
Threshold Synthesis Problem} 

\author{Jan-Georg Smaus\\
IRIT\\
Universit\'e Paul Sabatier Toulouse\\
France
\And
Christian Schilling\\
{\bf\Large Fabian Wenzelmann}\\
Institut f\"ur Informatik\\ 
Albert-Ludwigs-Universit\"at Freiburg\\
Germany}

\maketitle

\begin{abstract}
A \emph{linear pseudo-Boolean constraint} (LPB) is an expression of
the form $$a_1\cdot \ell_1+\ldots+a_m\cdot \ell_m\geq d,$$ where each
$\ell_i$ is a \emph{literal} (it assumes the value 1 or 0 depending on
whether a propositional variable $x_i$ is true or false) and
$a_1,\ldots,a_m,d$ are natural numbers. An LPB represents a Boolean
function, and those Boolean functions that can be represented by
exactly one LPB are called \emph{threshold functions}.  The problem of
finding an LPB representation of a Boolean function if possible is
called \emph{threshold recognition problem} or \emph{threshold
  synthesis problem}.  The problem has an $O(m^7t^5)$
algorithm using linear programming, where $m$ is the dimension and $t$
the number of terms in the DNF input. It has been an open question 
whether one can recognise threshold functions
through an entirely combinatorial procedure. Smaus has developed such a procedure for
doing this, which works by decomposing the DNF and ``counting'' the
variable occurrences in it. We have implemented
both algorithms as a thesis project. We report here on this
experience. The most important insight was that the algorithm by Smaus
is, unfortunately, incomplete.
\end{abstract}

\section{Introduction}\label{intro-sec}
A \emph{linear pseudo-Boolean constraint} (LPB)
\cite{DixGin-KER00,FraHer06} 
is an expression of the form
$a_1\ell_1+\ldots+a_m\ell_m\geq d$. Here each $\ell_i$ is a \emph{literal} of
the form $x_i$ or $\bar{x}_i\equiv1-x_i$, i.e.~$x_i$ becomes 0 if $x_i$
is false and 1 if $x_i$ is true, and vice versa for $\bar{x}_i$. 
Moreover, $a_1,\ldots,a_m,d$ are natural numbers. 

An LPB can be used to represent a Boolean\footnote{Whenever we say
  ``function'' we mean ``Boolean function''.}  function;
e.g.~$x_1+\bar{x}_2+x_3\geq3$ represents the same function as the
propositional formula $x_1\land\lnot x_2\land x_3$. It has been observed that a
function can be often represented more compactly as a set of LPBs than
as a \emph{conjunctive} or \emph{disjunctive} normal form (CNF or DNF)
\cite{DixGin-KER00,FraHer06}.  E.g.~the LPB
$2x_1+\bar{x}_2+x_3+x_4\geq2$ corresponds to the DNF $x_1\lor(\lnot
x_2\land x_3)\lor(\lnot x_2\land x_4)\lor(x_3\land x_4)$, 
which has four clauses. 

In this work we are concerned with functions that can be represented
by a single LPB, the so-called \emph{threshold functions}.  
The problem of recognising a Boolean function given in 
DNF as threshold function and computing the LPB representation 
if possible, is called \emph{threshold recognition problem} or
\emph{threshold synthesis problem}.  The problem is known to have an
$O(m^7t^5)$ algorithm using linear programming, where $m$ is the
dimension and $t$ the number of terms in the DNF \cite{CraHam07}. 

It has been an open question for decades whether it is possible to
recognise threshold functions through an entirely combinatorial
procedure, i.e., without resorting to the equivalent linear
program. Smaus has developed a procedure, which works
by decomposing the DNF and ``counting'' its variable occurrences 
in an appropriate way \cite{Sma-CPAIOR07}. 

Schilling and Wenzelmann, students of Freiburg University, have
implemented the classical linear programming algorithm and the more
recent combinatorial algorithm, respectively, as Bachelor thesis
projects \cite{SchillingBachelor11,WenzelmannBachelor11}. We report here on this experience. The most important
insight was that the algorithm by Smaus is, unfortunately, incomplete.

This paper is organised as follows. 
We continue with some preliminaries. Sec.~\ref{lp-sec} describes the
linear programming algorithm, Sec.~\ref{encoding-sec} the
combinatorial procedure, 
Sec.~\ref{impl-sec} the implementation, and Sec.~\ref{concl-sec} concludes and discusses future work. 
 
\section{Preliminaries}\label{prelim-sec}
We assume the reader to be familiar with the basic notions of
propositional logic. 

An \defining{$m$-dimensional Boolean function} $f$ is a function 
$\bool^m\to\bool$. 
A \defining{linear pseudo-Boolean constraint} (LPB)
is an inequality of the form
\begin{equation}\label{LPB-eq}
a_1\ell_1+\ldots+a_m\ell_m\geq d
\qquad a_i\in\nat, d\in\integ, \ell_i\in\{x_i,\bar{x}_i\}.
\end{equation}
We call the $a_i$ \defining{coefficients} and $d$ the
\defining{degree} \cite{Hoo92}. 
An occurrence of a \defining{literal} $x_i$ (resp., $\bar{x}_i$) is called an
occurrence of $x_i$ in \defining{positive} (resp., 
\defining{negative}) polarity. Note that if $d\leq0$, then the LPB is a
tautology. The reason for allowing for negative $d$ will become
apparent in Subsec.~\ref{decomposition-subsec}.

A \defining{DNF} is a 
formula of the form $c_1\lor\ldots\lor c_n$
where each \defining{clause} $c_j$ is a conjunction of literals. 
Formally,  a DNF is a set of sets of literals,
i.e., the order of clauses and the order of
literals within a clause are insignificant.
For DNFs, we
assume without loss of generality that no clause
is a subset of another clause
(the latter clause would be redundant since it is \emph{absorbed}).
We call a DNF \emph{prime irredundant} if every clause is a prime implicant, i.e.,
if for clause $c_1$ there is no clause $c_2 \neq c_1$ such that $c_1 \lor c_2 = c_2$.
Any Boolean function can be represented by a DNF \cite{Weg87-nohtml}.

It is easy to see that an LPB can only represent \emph{monotone}
functions, i.e., functions represented by a DNF where each variable
occurs in only one polarity. Hence any DNF containing a variable in
different polarities is immediately uninteresting for us. 
Without loss of generality, we assume that this polarity
is positive. 

\section{The linear programming algorithm}\label{lp-sec}
We shortly summarise the solution via linear programming, established by Peled \& Simeone \cite{PeledSimeone85,CraHam07}. 

For some DNFs, it is possible to establish a complete order $\succeq$ on the
variables which, intuitively speaking, has the following meaning:
$x_i\succeq x_j$ iff starting from any given input tuple $X^* \in \bool^m$,
setting $x_i^*$ to true is more likely to make the DNF true than
setting $x_j^*$ true. The functions represented by such a DNF are called 
\emph{regular}.

The algorithm first tests the input DNF for the
regularity property. The property is weaker than
the threshold property, and so if a DNF is not regular, then it is not
convertible and we must give up.

The order is established by counting the variables in a special way.
Intuitively, a variable is ``important'' if it occurs in \emph{many}
clauses and if it occurs in \emph{short} clauses. This is formalised
as the so-called \emph{occurrence pattern} of a variable $x$ in $\phi$,
written $\OP(\phi,x)$.  For space reasons, we do not give the formal
definition and refer the reader to \cite{Sma-CPAIOR07}.

Computing the set of occurrence patterns for all variables in $\phi$ can
be done in time linear in the size of $\phi$ as it can be done in a
single pass over $\phi$. In fact, the number of elements of all
occurrence patterns is exactly the number of literals in $\phi$.  Thus
sorting the variables w.r.t.~the occurrence patterns can be done in
time polynomial in $|\phi|$.

The notion of occurrence patterns is equivalent to the
so-called \emph{Winder matrix} \cite{winder62}. We will need the
concept again in the next section.

Provided the DNF is regular, 
we make use of the \emph{minimal true points} of the DNF, 
i.e.~the true tuples where we cannot set any 1-value to 0 without making the point false.
We also use the \emph{maximal false points} defined analogously.
Note that these together characterise the DNF uniquely.
In general, no polynomial algorithm is known to find these points
(which is no surprise since the general task is NP-complete \cite{PeledSimeone85}),
but for the special case that the input DNF is prime irredundant this is possible.
The reason is that the true points can be read directly from the clauses.
It is for this reason that we require the 
input DNF to be in prime irredundant form.

Having these, there exists a polynomial time procedure to find the maximal false points.
Then we can formulate the following linear program where the minimal true points are $x^1,\dots,x^k$ and the 
maximal false points are $y^1,\dots,y^l$:
\begin{alignat*}{8}
	\sum_{i=1}^m a_i x_i^j &\quad& \geq &\quad&& d &\quad& (1 \leq j \leq k) \\
	\sum_{i=1}^m a_i y_i^j && < &&& d && (1 \leq j \leq l) \\
	a_i && \geq &&& 0 && (1 \leq i \leq m)
\end{alignat*}
Note that the weights $a_i$ are the variables in the LP formulation and the threshold is $d$.
Finally, the linear program is passed to an LP solver.
The reason for the complexity blow-up ($O(m^7t^5)$ where $m$ is the dimension and $t$ the number of terms in the DNF)
is mainly due to the linear programming.
The other parts run in $O(m^2t)$, so the whole procedure gains from future improvements of linear programming.
It should be mentioned that for most inputs the well-known simplex method for solving linear programs runs in linear time.

\section{The combinatorial algorithm}\label{encoding-sec}
In this section we recall the results from our previous work
\cite{Sma-CPAIOR07} 
and present an algorithm for the problem of converting
a DNF to an equivalent LPB if possible.

\subsection{Determining the order of coefficients}\label{coefficient-order-subsec}
Given a DNF $\phi$, if $\phi$
can be represented as an LPB at all, then the coefficients must
respect the order $\succeq$ introduced in the previous section, i.e., 
$\OP(\phi,x_i)\succeq\OP(\phi,x_k)$
implies that $a_i\geq a_k$ in the resulting LPB:

\begin{lemma}\label{DNF-occurrence-pattern-lem}
Let $\phi$ be a DNF represented by the LPB $\coeffSum{1}{m}\geq d$. Then
$a_i\geq a_k$ implies $\OP(\phi,x_i)\succeq\OP(\phi,x_k)$; moreover, there exists an LPB 
$\coeffSumPrime{1}{m}\geq d'$ representing $\phi$ such that $\OP(\phi,x_i)=\OP(\phi,x_k)$
implies $a'_i=a'_k$. 
\end{lemma}

In our algorithm, one notion used is that of \emph{symmetry}: 
two variables in a DNF are symmetric if exchanging them yields the
same DNF. For space reasons, we neglect this aspect in the sequel and
refer the reader to \cite{Sma-CPAIOR07}.

\subsection{Decomposing a DNF}\label{decomposition-subsec}
We want to find an LPB
representing $\phi$ if possible. Using
Lemma \ref{DNF-occurrence-pattern-lem}, 
we can establish the order of the
coefficients. Assume the numbering of the variables
is such that we have $\OP(\phi,x_1)\succeq\ldots\succeq\OP(\phi,x_m)$. Consider now the
maximal set $X=\{x_1,\ldots,x_l\}$ such that $\OP(\phi,x_1)=\ldots=\OP(\phi,x_l)$ 
($=: \OP(\phi,X)$). 
(Of course, it is very well possible that $X=\{x_1\}$, i.e., $l=1$.)  
We want to divide $\phi$ into subproblems, and for this 
we partition $\phi$ according to how many
variables from $X$ each  clause contains. We then remove the variables
from $X$ from each clause, which gives $l+1$ subproblems (DNFs). Theorem
\ref{DNF-encoding-generalisation-thm} below states under which conditions 
solutions to these
subproblems can be combined to an LPB for $\phi$. However, since the
solutions have to be similar in a certain sense, it turns out that we
cannot simply solve the subproblems independently and \emph{then}
combine the solutions, but we must solve the subproblems in parallel,
as will be shown in Subsec.~\ref{composition-subsec}.

The following statements do not require $X$ to be \emph{maximal},
e.g.~if $\{x_1,\ldots,x_5\}$ is the maximal set such that
$\OP(\phi,x_1)=\ldots=\OP(\phi,x_5)$, then the statements will also hold for
$X=\{x_1,x_2,x_3\}$. From now on, the letter $X$ will always denote 
a set as just described, maximal or not.

\begin{definition}\label{Cut-def}
Let $\phi$ be a DNF and $X$ a subset of its variables with $|X| =l$.
If $\phi$ contains a clause $c\subseteq X$, then let $k_{\max}$ be the length of
the longest such clause; otherwise let $k_{\max}:=\infty$. 
For $0\leq k\leq l$,  we define
$\Cut{\phi}{X}{k}$ as the disjunction of  clauses from $\phi$ 
containing exactly $\min\{k,k_{\max}\}$ variables from $X$, with those 
variables removed. 
\end{definition}

When constructing the $\Cut{\phi}{X}{k}$ from $\phi$, we say that we \emph{split away} 
the variables in $X$ from $\phi$. 

\begin{example}\label{Cut-ex}
Let $\phi\equiv(x_1)\lor(x_2)\lor(x_3\land x_4)$ and $X=\{x_1,x_2\}$.
We have $k_{\max}=1$.
Then $\Cut{\phi}{X}{0}=(x_3\land x_4)$, 
$\Cut{\phi}{X}{1}=\true$ (i.e.~the disjunction of twice the 
empty conjunction), and 
$\Cut{\phi}{X}{2}=\true$.
\end{example}

We must solve the $l+1$ subproblems in such a way that the
resulting LPBs agree in all coefficients, and that the degree difference of
neighbouring LPBs is always the same. Before giving the theorem,
we give two examples for illustration. 

\begin{example}\label{simplest-alg-ex}
Consider 
$\phi\equiv(x_1\land x_2)\lor(x_1\land x_3)\lor(x_1\land x_4)\lor(x_2\land x_3\land x_4)$ and $X=\{x_1\}$.
Then $\Cut{\phi}{X}{0}=x_2\land x_3\land x_4$, represented by $x_2+x_3+x_4\geq3$.
Moreover, $\Cut{\phi}{X}{1}=x_2\lor x_3\lor x_4$, represented by $x_2+x_3+x_4\geq1$.

Since the coefficients of the two LPBs agree, it turns out that $\phi$
can be represented by $2x_1+x_2+x_3+x_4\geq3$. The coefficient of $x_1$
is given by the difference of the two degrees, i.e.~$3-1$.
\end{example}

\begin{example}\label{LPB-adaptation-ex}
Consider 
$\phi\equiv(x_1\land x_2)\lor(x_1\land x_3\land x_4)\lor(x_2\land x_3\land x_4)$ and
$X=\{x_1,x_2\}$. We have
$\Cut{\phi}{X}{0}=\false$, represented by $x_3+x_4\geq4$, 
$\Cut{\phi}{X}{1}=x_3\land x_4$, represented by $x_3+x_4\geq2$, and  
$\Cut{\phi}{X}{2}=\true$, represented by $x_3+x_4\geq0$. 
The DNF $\phi$ is represented by $2x_1+2x_2+x_3+x_4\geq4$. 
The coefficient of $x_1,x_2$ is given by $4-2=2-0=2$ 
(the degrees are ``equidistant'').   
\end{example}

\begin{theorem}\label{DNF-encoding-generalisation-thm}
Let $\phi$ be a DNF in variables $x_1,\ldots,x_m$ and suppose 
$X=\{x_1,\ldots,x_l\}$ are symmetric variables such that 
$\OP(\phi,X)$ is maximal 
w.r.t.~$\preceq$ in $\phi$. Then $\phi$ is represented by an LPB 
$\coeffSum{1}{m}\geq d$, where $a_1=\ldots=a_l$, iff for all $k\in\zerotol$,
the DNF $\Cut{\phi}{X}{k}$ is represented by 
$\coeffSum{l+1}{m}\geq d-k\cdot a_1$. 

\addtocounter{equation}{1}
\end{theorem}

The remaining problem is that a DNF might be represented by various LPBs, and so
even if the LPBs computed recursively do not have agreeing coefficients and
equidistant degrees, one might find alternative LPBs 
(such as the non-obvious LPB for $\false$ in
Ex.~\ref{LPB-adaptation-ex}) so that 
Thm.~\ref{DNF-encoding-generalisation-thm} can be applied.

Before addressing this problem, we generalise LPBs by recording to what extent
degrees can be shifted without changing the meaning. To formulate this, we
temporarily lift the restriction that coefficients and degrees must be
integers. How to obtain integers in the end is explained at the end of
Subsec.~\ref{composition-subsec}.

\begin{definition}
Given an LPB $I\equiv \coeffSum{1}{m}\geq d$, we call 
$\mint$ the \defining{minimum degree} of $I$ if 
$\mint$ is the smallest number (possibly $-\infty$) such that for any 
$\mint'\in (\mint,d]$,  the LPB
$\coeffSum{1}{m}\geq \mint'$ represents the same  function as $I$. 
We call $\maxt$ the \defining{maximum degree} if $\maxt$ is the 
biggest number (possibly $\infty$) such that 
$\coeffSum{1}{m}\geq \maxt$ represents the same  function as $I$.
\end{definition}

Note that the minimum degree of $I$ is itself not a possible degree
of $I$.
Since the minimum and maximum degrees of an LPB are more
informative than its actual degree, we introduce the
notation $\coeffSum{1}{m}\geq \degree{\mint}{\maxt}$ for denoting an LPB with minimum
degree $\mint$ and maximum degree $\maxt$.

The next lemma strengthens Thm.~\ref{DNF-encoding-generalisation-thm}, stating
that information about minimum and maximum degrees can be maintained with
little overhead.

\begin{lemma}\label{degree-slack-lem}
Make the same assumptions as in
Thm.~\ref{DNF-encoding-generalisation-thm}, and assume that  
for all $k\in\zerotol$, the DNF $\Cut{\phi}{X}{k}$ is represented by 
$I^k\equiv \coeffSum{l+1}{m}\geq d-k\cdot a_1$. Moreover, for all
$k\in\zerotol$, let $\mint_k,\maxt_k$ be minimum and maximum
degrees of $I^k$, respectively. Then
$\mint:=\max_{k\in\zerotol}(\mint_k+k\cdot a_1)$, 
$\maxt:=\min_{k\in\zerotol}(\maxt_k+k\cdot a_k)$ are the 
minimum and maximum degrees of $\coeffSum{1}{m}\geq d$. 
\end{lemma}

\addtocounter{equation}{1}

\subsection{Composing LPBs}\label{composition-subsec}
Theorem \ref{DNF-encoding-generalisation-thm} suggests a recursive algorithm
where, at least conceptually, in the base case we have at most $2^{m}$ trivial
problems of determining an LPB, trivial since the formula for which we must find
an LPB is either $\true$ or $\false$.

\begin{table*}[t]
\begin{center}
\begin{tabular}{l|l|l|l|l|l}
&                      &                      &                             & \ld{$\CutT{x_{3..4}}{0}\equiv f$} & $\CutT{x_{3..5}}{0}\equiv f$\\
&$\CutT{x_1}{0}$        & $\CutT{x_2}{0}\equiv$      & \sd{$\CutT{x_{3}}{0}\equiv f$}      & \ld{$\CutT{x_{3..4}}{1}\equiv f$} & $\CutT{x_{3..5}}{1}\equiv f$ \\
&$\equiv(x_2\land x_3)\lor$         & $(x_3\land x_4\land x_5)$      & \sd{$\CutT{x_{3}}{1}$}         &  \ld{$\CutT{x_{3..4}}{2}\equiv x_5$} & $\CutT{x_{3..5}}{2}\equiv f$ \\
&$(x_2\land x_4)\lor$          &                      & \sd{$\equiv (x_4\land x_5)$}            &                         &  $\CutT{x_{3..5}}{3}\equiv t$\\\cline{4-6}
&$(x_3\land x_4\land x_5)$      & $\CutT{x_2}{1}\equiv$    & \ld{$\CutT{x_{3}}{0}\equiv x_4$}      & $\CutT{x_{3..4}}{0}\equiv f$ & \\
&                     &    $x_3\lor x_4$         & \ld{$\CutT{x_{3}}{1}\equiv t$}       & $\CutT{x_{3..4}}{1}\equiv t$ & \\
\raisebox{-1ex}[0cm][0cm]{$\phi$}&     &     &                                 & $\CutT{x_{3..4}}{2}\equiv t$ & \\\cline{3-6}
&&&&&\\[-1.7ex]                   
&                     &                    & \sd{$\CutT{x_{2..3}}{0}\equiv$}         & \ld{$\CutT{x_{2..4}}{0}\equiv x_5$} &$\CutT{x_{2..5}}{0}\equiv f$ \\
&$\CutT{x_1}{1}$       &\ud{$\CutT{x_2}{0}\equiv$}& \sd{$x_4\lor x_5$}               & \ld{$\CutT{x_{2..4}}{1}\equiv t$} &$\CutT{x_{2..5}}{1}\equiv t$ \\
&$\equiv x_2\lor x_3$            &\ud{$x_3\lor x_4\lor x_5$}     & \sd{$\CutT{x_{2..3}}{1}\equiv t$}       & \ld{$\CutT{x_{2..4}}{2}\equiv t$} &$\CutT{x_{2..5}}{2}\equiv t$ \\
&$\lor x_4\lor x_5$            &\ud{$\CutT{x_2}{1}\equiv t$}& \sd{$\CutT{x_{2..3}}{2}\equiv t$}        & \ld{$\CutT{x_{2..4}}{3}\equiv t$} & $\CutT{x_{2..5}}{3}\equiv t$\\
&                     &                   &                                &                    &$\CutT{x_{2..5}}{4}\equiv t$
\end{tabular}\caption{The recursive problems of Ex.~\ref{op-cont-ex}\label{op-cont-table}}
\vspace{-3ex}
\end{center}
\end{table*}

\begin{example}\label{op-cont-ex}
Consider $\phi\equiv(x_1\land x_2)\lor(x_1\land
x_3)\lor(x_1\land x_4)\lor(x_1\land x_5)\lor(x_2\land x_3)\lor(x_2\land x_4)\lor(x_3\land x_4\land x_5)$.
To find an LPB for $\phi$, we must find LPBs for
$\Cut{\phi}{\{x_1\}}{0}$ 
and $\Cut{\phi}{\{x_1\}}{1}$.
To find an LPB for $\Cut{\phi}{\{x_1\}}{0}$, we must find LPBs for
$\Cut{\Cut{\phi}{\{x_1\}}{0}}{\{x_2\}}{0}$ and 
$\Cut{\Cut{\phi}{\{x_1\}}{0}}{\{x_2\}}{1}$, and so forth. Table \ref{op-cont-table} gives all 
the formulae for which we must find LPBs. 
For a concise notation we use some abbreviations which we explain using 
$\CutT{x_{3..5}}{0}\equiv f$ in the top-right corner: it stands for  
$\Cut{(x_3\land x_4\land x_5)}{\{x_3,x_4,x_5\}}{0}\equiv\false$, i.e.~the 
`$\cdot$' stands for the nearest \emph{non-shaded} formula to the left, here 
$(x_3\land x_4\land x_5)$. Note how we arranged the subproblem formulae in the table: 
e.g.~$(x_3\land x_4\land x_5)$ has \emph{three} symmetric variables that are split away to
obtain the subproblems to be solved, so these subproblems are located \emph{three}
columns to the right of $(x_3\land x_4\land x_5)$. The two shaded boxes in between contain the
subproblems obtained by splitting away only $\{x_3\}$, $\{x_3,x_4\}$, resp.
Observe also the empty box in the last column, arising from the fact that we do
not attempt to split away $x_5$ from $x_3\lor x_4$. 
\end{example}

The algorithm we propose is not a purely recursive one, since the subproblems
at each level must be solved in parallel. 
Explained using the example, we first find LPBs for the
formulae in the rightmost column, which have $0$ variables and hence we must
determine $0$ coefficients. Next to the left, we have formulae that contain (at
most) $x_5$, and we determine LPBs representing these, where we use the same
$a_5$ for all formulae! Then we determine $a_4$, and so forth.

Taking $(x_3\land x_4\land x_5)$ in Table \ref{op-cont-table} as an example,
Thm.~\ref{DNF-encoding-generalisation-thm} suggests that $a_3,a_4,a_5$ should be
equal ($x_3,x_4,x_5$ are symmetric) and determined in one go.  However,
since $a_3,a_4,a_5$ also have to represent other subproblem formulae where
$x_3,x_4,x_5$ are not necessarily symmetric, one cannot determine $a_3,a_4,a_5$
in one go, but rather first $a_5$, then $a_4$, then $a_3$. Therefore, it is
necessary to define and interpret formulae obtained by splitting away fewer
variables than one could split away, in the sense of
Thm.~\ref{DNF-encoding-generalisation-thm}. These are the shaded formulae.

For each $l\in\{0,\ldots m\}$, we call the formulae in column $l+1$ the $l$-\emph{successors}. Shaded formulae are
called \emph{auxiliary}, the others are called \emph{main}. Formulae that have
no further formulae to the right are called \emph{final}. The following
definition formalises these notions. 

\begin{definition}\label{successor-def}
Let $\phi$ be a DNF in $m$ variables. Then $\phi$ is the $0$-\defining{successor} of
$\phi$. Furthermore, $\phi$ is a \defining{main} successor of $\phi$. 
Moreover, if $\phi'$ is a main $n$-successor of $\phi$, and $l$ is maximal 
so that $x_{n+1},\ldots,x_{n+l}$ are  symmetric in
$\phi'$, then for all $l', k$ with $1\leq l'\leq l$ and 
$0\leq k\leq l'$, we say that $\Cut{\phi'}{\{x_{n+1},\ldots,x_{n+l'}\}}{k}$ is an $(n+l')$-successor of
$\phi$. The  $(n+l)$-successors are called \defining{main}, and for $l'<l$, the 
$(n+l')$-successors are called \defining{auxiliary}. 
A node that is a main node and $\true$ or $\false$ is 
called \emph{final}. 
\end{definition}

Note in particular $x_3\lor x_4$ in column 3 in Table \ref{op-cont-table}. It does
not contain $x_5$, and so we obtain final $4$-successors in the
last-\emph{but-one} column. Clearly, a final successor of $\phi$ is either $\true$
or $\false$.

\begin{table*}[t]
\begin{center}
\begin{tabular}{r|r|r|r|r|r}
$4x_1+3x_2+$&     $3x_2+$  &                      &                          &                    &              \\
$2x_3+2x_4+$&$2x_3+2x_4+$  & $2x_3+2x_4+$            &        $2x_4+$            &                      & $\sum_{i=6}^5a_ix_i$ \\
$x_5\geq\ldots$&$x_5\geq\ldots$        & $x_5\geq\ldots$               &  $x_5\geq\ldots$               &   $x_5\geq\ldots$             &  $\geq\ldots$    \\\hline\hline
&                      &                      &                             & \ld{$\LPBT{x_{3..4}}{0}\degree{1}{\infty}$} & $\LPBT{x_{3..5}}{0}\degree{0}{\infty}$\\
&$\LPBT{x_1}{0}$        & $\LPBT{x_2}{0}$      & \sd{$\LPBT{x_{3}}{0}\degree{3}{\infty}$}     & \ld{$\LPBT{x_{3..4}}{1}\degree{1}{\infty}$} & $\LPBT{x_{3..5}}{1}\degree{0}{\infty}$ \\
&                     & \ub{$\degree{4}{5}$}         & \sd{$\LPBT{x_{3}}{1}\degree{2}{3}$}      &  \ld{$\LPBT{x_{3..4}}{2}\degree{0}{1}$} & $\LPBT{x_{3..5}}{2}\degree{0}{\infty}$ \\
&$\degree{4}{5}$              &                      &                                 &                         &  $\LPBT{x_{3..5}}{3}\degree{-\infty}{0}$\\\cline{4-6}
&                     & $\LPBT{x_2}{1}$    & \ld{$\LPBT{x_{3}}{0}\degree{1}{2}$}      & $\LPBT{x_{3..4}}{0}\degree{1}{\infty}$ & \\
&                     &    $\degree{1}{2}$         & \ld{$\LPBT{x_{3}}{1}\degree{-\infty}{0}$}       & $\LPBT{x_{3..4}}{1}\degree{-\infty}{0}$ & \\
\raisebox{-1ex}[0cm][0cm]{$\degree{4}{5}$}&     &     &                                 & $\LPBT{x_{3..4}}{2}\degree{-\infty}{0}$ & \\\cline{3-6}
&&&&&\\[-2.7ex]                   
&                     &                    &                                 & \ld{$\LPBT{x_{3..4}}{0}\degree{0}{1}$} &$\LPBT{x_{3..5}}{0}\degree{0}{\infty}$ \\
&$\LPBT{x_1}{1}$       &                    &\sd{$\LPBT{x_{3}}{0}\degree{0}{1}$}      & \ld{$\LPBT{x_{3..4}}{1}\degree{-\infty}{0}$} &$\LPBT{x_{3..5}}{1}\degree{-\infty}{0}$ \\
&$\degree{0}{1}$              &\ud{$\LPBT{x_2}{0}\degree{0}{1}$}& \sd{$\LPBT{x_{3}}{1}\degree{-\infty}{0}$}       & \ld{$\LPBT{x_{3..4}}{2}\degree{-\infty}{0}$} &$\LPBT{x_{3..5}}{2}\degree{-\infty}{0}$ \\
&                     &\ud{$\LPBT{x_2}{1}\degree{-\infty}{0}$}& \sd{$\LPBT{x_{3}}{2}\degree{-\infty}{0}$}   & \ld{$\LPBT{x_{3..4}}{3}\degree{-\infty}{0}$} & $\LPBT{x_{3..5}}{3}\degree{-\infty}{0}$\\
&                     &                   &                                &                    &$\LPBT{x_{3..5}}{4}\degree{-\infty}{0}$
\end{tabular}
\end{center}
\caption{LPBs for Ex.~\ref{op-cont-ex}\label{op-cont-thresh-table}}
\vspace{-3ex}
\end{table*}

\begin{proposition}\label{split-away-fewer-prop}
Assume $\phi$, $\phi'$, $n$, $l$ as in Def.~\ref{successor-def}. For  
$0< l'<l$ and $0\leq k\leq l'$, we have 
\[
\begin{array}{l}
\Cut{\Cut{\phi'}{\{x_{n+1},\ldots,x_{n+l'}\}}{k}}{\{x_{n+l'+1}\}}{0}\equiv
\\\quad\Cut{\phi'}{\{x_{n+1},\ldots,x_{n+l'+1}\}}{k}\\
\Cut{\Cut{\phi'}{\{x_{n+1},\ldots,x_{n+l'}\}}{k}}{\{x_{n+l'+1}\}}{1}\equiv
\\\quad\Cut{\phi'}{\{x_{n+1},\ldots,x_{n+l'+1}\}}{k+1}
\end{array}
\]
\end{proposition}

For example, consider $\Cut{(x_3\land x_4\land x_5)}{\{x_3\}}{1}\equiv (x_4\land x_5)$ in Table
\ref{op-cont-table}. We have 
$\Cut{(x_4\land x_5)}{\{x_4\}}{0}\equiv\Cut{(x_3\land x_4\land x_5)}{\{x_3,x_4\}}{1}$ and 
$\Cut{(x_4\land x_5)}{\{x_4\}}{1}\equiv\Cut{(x_3\land x_4\land x_5)}{\{x_3,x_4\}}{2}$. 
Generally, each non-final successor is associated with two formulae in the 
column right next to it, one slightly up and one slightly down, obtained by 
splitting away the variable with the smallest index. 

This is not surprising per se 
and corresponds to a na\"\i ve approach where we 
always 
split away one variable at a time 
(for applying Thm.~\ref{DNF-encoding-generalisation-thm}), thereby
constructing $2^m$ formulae in the rightmost column. The point of 
Prop.~\ref{split-away-fewer-prop} is that 
we can usually construct fewer formulae since 
$\Cut{\Cut{\phi}{\{x_{n+1},\ldots,x_{n+l'}\}}{k}}{\{x_{n+l'+1}\}}{1}$
and
$\Cut{\Cut{\phi}{\{x_{n+1},\ldots,x_{n+l'}\}}{k+1}}{\{x_{n+l'+1}\}}{0}$ 
coincide. 
This means, $\phi'$ triggers $l+1$ main $(n+l)$-successors instead of $2^l$.
In Table \ref{op-cont-table}, we have 12 final
formulae rather than $2^5=32$.

\label{only12}

It seems to be generally the case that the table has much fewer final
nodes that $2^m$.  The many examples we looked at strongly suggest
that even if one tries to construct an input DNF that has as few
symmetries as possible and hence would lead to a big table, the
subformulae constructed by the splitting always exhibit many
symmetries. It would be interesting to have a theoretical statement
about this observation. 

The following theorem states if and how one can find the next coefficient and
degrees for representing all $k$-successors of $\phi$ provided one has
coefficients and degrees for representing all $(k+1)$-successors.

\begin{theorem}\label{ak-completion-criterion-thm}
Assume $\phi$ as in Thm.~\ref{DNF-encoding-generalisation-thm} and some $k$ with 
$0\leq k\leq m-1$, and let $\Phi_k$ be the set of $k$-successors of $\phi$.
For every non-final $\phi'\in\Phi_k$, suppose we have two LPBs
$\coeffSum{k+2}{m}\geq\degree{\mint_{\phi'0}}{\maxt_{\phi'0}}$ and 
$\coeffSum{k+2}{m}\geq\degree{\mint_{\phi'1}}{\maxt_{\phi'1}}$, representing $\Cut{\phi'}{\{x_{k+1}\}}{0}$ and
$\Cut{\phi'}{\{x_{k+1}\}}{1}$, respectively. 

If it is possible to choose $a_{k+1}$ such that 
\begin{equation}\label{maxmin-diff-eq}
\max_{\phi'\in\Phi_k} (\mint_{\phi'0}-\maxt_{\phi'1}) < a_{k+1}< \min_{\phi'\in\Phi_k} (\maxt_{\phi'0}-\mint_{\phi'1}),
\end{equation}
then for all $\phi'\in\Phi_k$, the LPB $\coeffSum{k+1}{m}\geq\degree{\mint_{\phi'}}{\maxt_{\phi'}}$ represents
$\phi'$, where
\begin{gather}
\mint_{\phi'}=\max\{\mint_{\phi'0},\mint_{\phi'1}+a_{k+1}\},\;\notag \\ 
\!\!\maxt_{\phi'}=\min\{\maxt_{\phi'0},\maxt_{\phi'1}+a_{k+1}\} \ \mbox{for non-final $\phi'$};\label{new-minmax-eq}\\
\mint_{\phi'}=-\infty,\;\maxt_{\phi'}=0 \ \mbox{for $\phi'\equiv\true$};\\ 
\!\!
\mint_{\phi'}={\textstyle\sum_{i=k+1}^m}a_i,\;\maxt_{\phi'}=\infty \ \mbox{for $\phi'\equiv\false$.}\label{new-minmax-2-eq}
\end{gather}
If $\max_{\phi'\in\Phi_k} (\mint_{\phi'0}-\maxt_{\phi'1}) \geq \min_{\phi'\in\Phi_k} (\maxt_{\phi'0}-\mint_{\phi'1})$, 
then no $a_{k+1}$, $\mint_{\phi'}$,
$\maxt_{\phi'}$ exist such that 
$\coeffSum{k+1}{m}\geq\degree{\mint_{\phi'}}{\maxt_{\phi'}}$ represents $\phi'$ for all $\phi'\in\Phi_k$.

\end{theorem}

The $m$-successors of $\phi$, i.e., the formulae in the rightmost column,
can only be $\false$ or $\true$. 
They are represented by LPBs with an empty sum as
l.h.s.: $\sum_{i=m+1}^ma_ix_i\geq\degree{0}{\infty}$ for $\false$,
$\sum_{i=m+1}^ma_ix_i\geq\degree{-\infty}{0}$ for $\true$. Then we proceed using
Thm.~\ref{ak-completion-criterion-thm}, in each step choosing an
arbitrary $a_{k+1}$ fulfilling (\ref{maxmin-diff-eq}).

\begin{example}\label{op-cont2-ex}
Consider again Ex.~\ref{op-cont-ex}. Table
\ref{op-cont-thresh-table} is arranged in strict correspondence to Table
\ref{op-cont-table} and shows LPBs for all successors of $\phi$. 
In the top line we give the l.h.s.~of the LPBs, which is of
course the same for each LPB in a column.
In the main table, we list the minimum and maximum degree of each formula. 

In the first step, applying (\ref{maxmin-diff-eq}), we have to choose $a_5$
so that
\[
\begin{array}{l}
\max\{0-\infty,0-\infty,0-0, 0-0,-\infty-0,-\infty-0,\\
\ \ \ -\infty-0\}<a_5<\min\{\infty-0,\infty-0, \infty--\infty,\\
\multicolumn{1}{r}{\infty--\infty,0--\infty,0--\infty,0--\infty\}.}
\end{array}
\]  
Choosing $a_5=1$ will do. The minimum and maximum degrees in column 5
are computed using (\ref{new-minmax-eq}); e.g.~the topmost $\degree{1}{\infty}$ is 
$\degree{\max\{0,0+1\}}{\min\{\infty,\infty+1\}}$. 

In the next step, we have to choose $a_4$ so that 
 \[
\begin{array}{l}
\max\{1-\infty,1-1, 1-0,-\infty-0,  0-0,-\infty-0,\\
\ \ \ -\infty-0\}<a_4<\min\{\infty-1,\infty-0, \infty--\infty,\\
\multicolumn{1}{r}{0--\infty,1--\infty,0--\infty,0--\infty\}.}
\end{array}
\]  
Choosing $a_4=2$ will do. Note that the bound $1-0<a_4$ comes from the
middle box of the fifth column and thus ultimately from $x_3\lor x_4$. Our
algorithm enforces that $a_4>a_5$, which must hold for an LPB representing 
$x_3\lor x_4$.

In the next step, $a_3$ can also be
chosen to be any number $>1$ so we choose $2$ again.
In the next step, $2<a_2<4$
must hold so we choose $a_2=3$. Finally, $3<a_1<5$ must hold so we choose
$a_1=4$. We obtain the LPB $4x_1+3x_2+2x_3+2x_4+x_5\geq\degree{4}{5}$.
\end{example}

\begin{table*}[t]
\begin{center}
\includegraphics[scale=1]{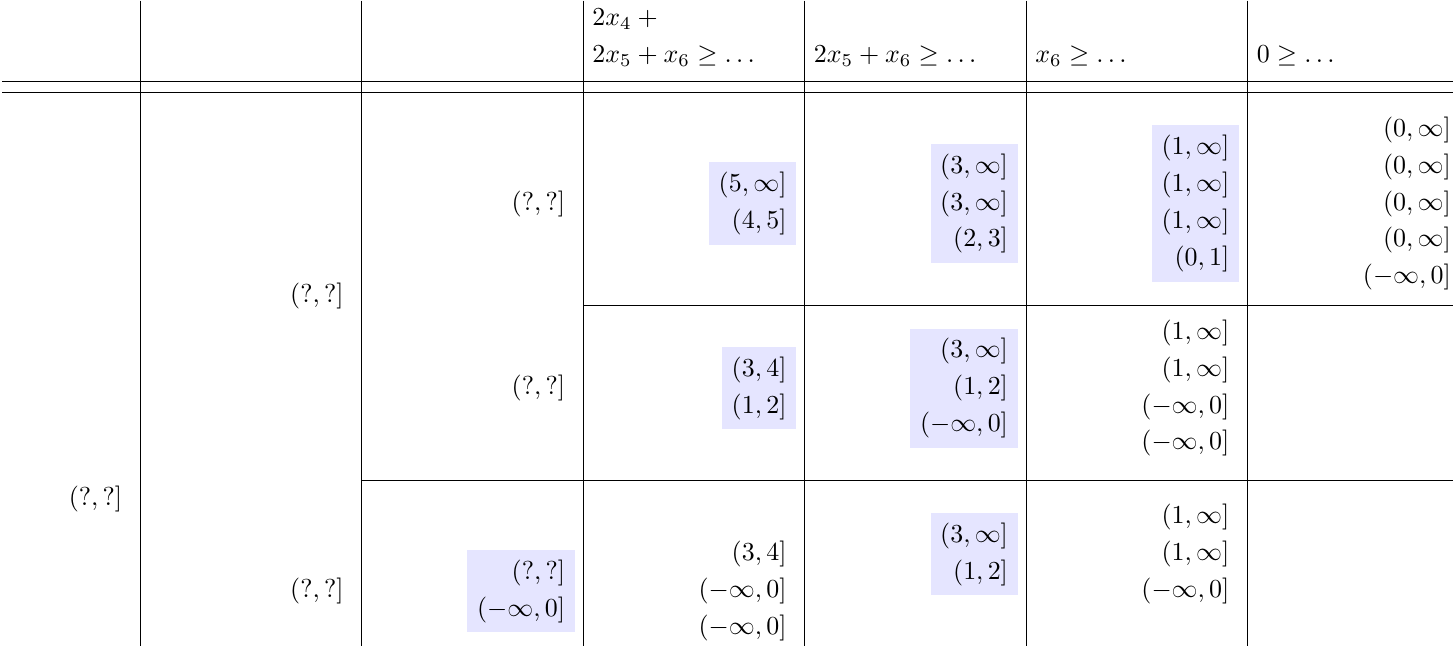}
\end{center}
\caption{LPBs for Ex. \ref{incompleteness-example}}
\label{table-incompleteness-solutions}
\end{table*}

We have seen in the example how our algorithm works. However, since the choice
of $a_{k+1}$ is not unique in general, one might be worried that a bad choice
of $a_{k+1}$ might later lead to non-applicability of
Thm.~\ref{ak-completion-criterion-thm}.
Contrary to what was stated by Smaus \shortcite{SmaTR230-07nociteother},
this is indeed a problem. We have suggested
to choose $a_{k+1}$ always as the smallest possible integer value to obtain an LPB
with small coefficients.
But it turns out that this strategy sometimes leads to a dead end.

\begin{example}\label{incompleteness-example}
Consider the DNF
\begin{align*}
\phi \equiv \left( x_1 \land x_2 \right) \lor \left( x_1 \land x_3 \right) \lor \left(x_1 \land x_4 \land x_5\right) \left(x_2 \land x_3 \land x_4\right)\\
\lor \left(x_2 \land x_3 \land x_5 \right) \lor \left( x_2 \land x_4 \land x_5 \right) \lor \left( x_3 \land x_4 \land x_5 \land x_6 \right)
\end{align*}
We apply the algorithm to create all successors of $\phi$ and calculate
LPBs for all recursive subproblems. The corresponding LPBs can be
found in Table \ref{table-incompleteness-solutions}. By applying the
strategy of choosing the coefficient as small as possible we choose 
$a_6 = 1, a_5=2, a_4 = 2$. We use the minimum and maximum degrees in
the fourth column to choose the coefficient $a_3$. We have to choose $a_3$ s.\,t.
\begin{align*}
&\max\left\lbrace 5-5,\, 3-2\,, 3-0,\, -\infty - 0 \right\rbrace < a_3 <\\
&\min\left\lbrace \infty - 4,\, 4-1,\, 4 -- \infty,\, 0 -- \infty \right\rbrace
\end{align*}
i.\,e. $3 < a_3 < 3$. This is, of course, not possible. But $\phi$ can be represented by the LPB $9x_1 + 7x_2 + 6x_3 + 4x_4 + 4x_5 + x_6 \geq 15$.

The algorithm found solutions for all subproblems in the fourth
column. 
But we cannot combine the coefficients chosen so far to a solution representing all LPBs in the third column.

Alternatively, we were allowed to choose $a_5 = a_4 = 4$, and if we do so, we obtain an appropriate LPB. Therefore the applicability of Thm.~\ref{ak-completion-criterion-thm} depends on the choice of the previous coefficients.
\end{example}

Another problem seems to be that $a_{k+1}$ could be forced to be
between neighbouring integers, in which case it cannot be an integer
itself. However, 
in this case, one can multiply all LPBs of the current system
by 2 (this obviously preserves the meaning of the LPBs) before
proceeding so that $a_{k+1}$ can be chosen to be an integer.

From the construction of the successors (see Table
\ref{op-cont-table}) it follows that all formulae 
in a column together have size less than all formulae 
in the column to the left of it, so that 
the entire table has size less than $|\phi|\cdot (m+1)$. One can thus show
that the complexity of the algorithm is polynomial in the size of $\phi$,
while the size of $\phi$ itself can be exponential in $m$. 
In fact, this is the most interesting case, because
in this case an LPB representation may yield an
exponential saving.

\section{Implementation}\label{impl-sec}
Both algorithms have been implemented in Java. They share the same core classes representing the main components such as DNFs and LPBs. The linear program is solved by \verb|lp_solve|.
Both implementations can be accessed and tested via a graphical user interface.

For testing the implementation we generated a full enumeration of LPBs up to seven variables. For LPBs with more variables we tested 180,000 randomly generated LPBs (with 8 to 25 variables). We transformed the LPBs to DNFs (so we know that for these DNFs there exists an LPB) to test the implementations. As expected the linear programming algorithm solved all tested input DNFs.

\begin{figure}
\includegraphics[scale=1]{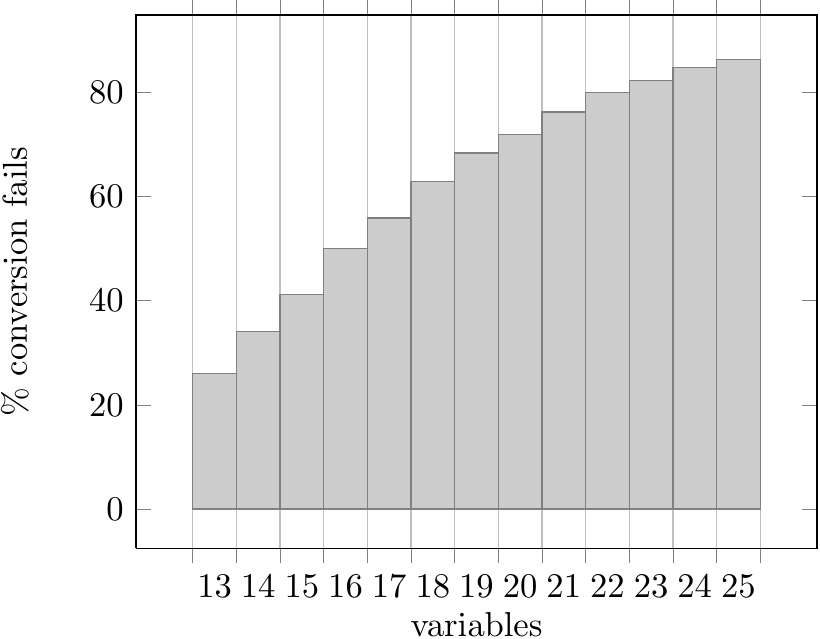}
\caption{Failure rate of the combinatorial algorithm}
\label{fig:failure}
\end{figure}

The combinatorial algorithm was able to solve all input DNFs with up
to five variables. But it fails on some DNFs with six variables (with the strategy to choose $a_{k+1}$ as small as possible). Our empirical analysis shows that the more variables a DNF contains the more often the conversion fails. Circa 8\% of the tested DNFs with seven variables cannot be converted, for DNFs with 25 variables circa 86\% cannot be converted. The failure rate for 13 to 25 variables is illustrated in Figure \ref{fig:failure}.

The linear programming algorithm was faster in direct runtime comparison, but we're still working on improvements for the combinatorial algorithm.

\begin{figure*}
\begin{center}
\includegraphics[scale=1]{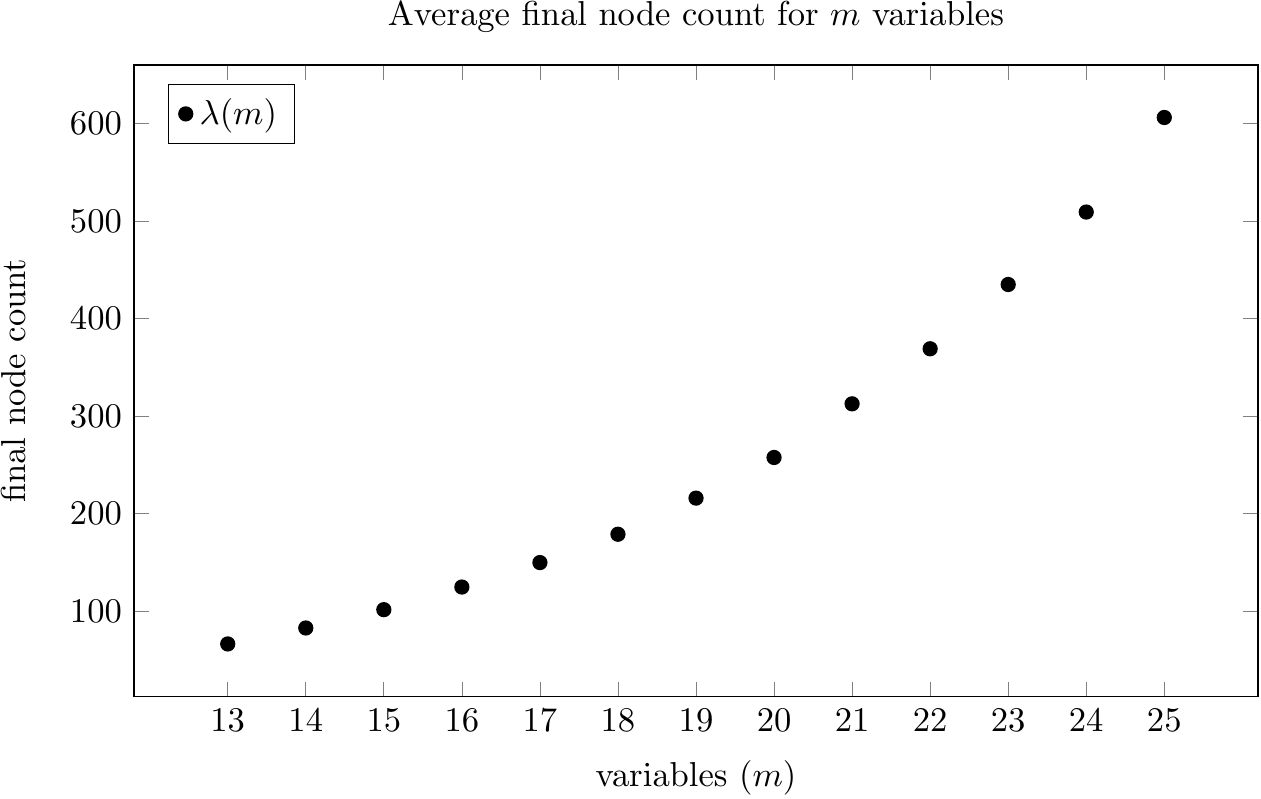}
\caption{$\lambda(m)$ is the average number of final nodes for DNFs with $m$ variables.}
\label{fig:final}
\end{center}
\end{figure*}

As discussed in Subsec.~\ref{only12}, 
for the combinatorial algorithm the number of \emph{final nodes} is an
important criterion for its theoretical runtime. Figure \ref{fig:final}
gives a first impression. It is hard to judge whether the growth
exhibited is exponential, but in any case, the number of final nodes
is much smaller than $2^m$: around 50000 times smaller for $m=25$. 

\section{Conclusion and future work}\label{concl-sec}
Linear pseudo-Boolean constraints have attracted interest because they
can often be used to represent Boolean functions more compactly than
CNFs or DNFs, and because techniques applied in CNF-based
propositional satisfiability solving can be generalised to LPBs
\cite{DixGin-KER00,FraHer06}. 

Some Boolean functions can be represented by a single LPB. 
The problem of
finding this LPB representation is
called \emph{threshold recognition problem}.  In this work, we have
implemented two algorithms for this problem, a classical one based on
linear programming, and a more recent one that we have previously
presented. The most important insight was that our algorithm 
is, unfortunately, incomplete. 

The most important topic for future work is, of course, trying to
reestablish completeness.

The obvious way to achieve this is to incorporate some kind of
backtracking into the algorithm: If a DNF can be represented by an LPB
and we cannot choose $a_{k+1}$, then this is because we must have
chosen one of the coefficients $a_{k+2},\ldots,a_m$ too small, because our
strategy so far was to chose the coefficients as small as possible. In
order to find a solution we increment the coefficients $a_{k+2},\ldots,a_m$
and re-evaluate the LPBs. We can use the minimum and maximum degrees
to ensure that we enumerate only legal candidates. We iteratively
increment the coefficients until we can choose the coefficient
$a_{k+1}$.

One problem of this approach is that for a DNF that cannot be
represented by an LPB, termination is not guaranteed, because
frequently the choice of the next coefficient is not bounded from
above. However, we are confident that this problem can be resolved
because it should be possible to derive some upper bound for each
variable in the sense that it is never necessary to choose a
coefficient bigger than this bound (something along the lines: it is
never necessary to choose a coefficient more than $m$ times bigger
than the previous coefficient). 

The other problem is of course that backtracking worsens that runtime
of the algorithm, and we very much fear that it will destroy the
polynomial runtime of the algorithm. 

The backtracking approach has been implemented and was able to find a
solution for each tested input DNF. But the implementation has also
shown that the higher the dimension $m$, the more often we have to use
backtracking. 

Alternatively, or more likely, additionally, one might use the
occurrence patterns for estimating the weight ratio:
 In the example above we were able to represent all LPBs
  in the fourth column but we were not able to choose $a_3$ such that we
  can represent all LPBs in the third column with the configuration
  $a_6 = 1, a_5 = a_4 = 2$. We need some \emph{global} information
  that the distance between $a_6$ and $a_5, a_4$ will be too small in
  the sequel.

  Maybe it is possible to use the occurrence patterns to formulate
  such constraints, i.e., one might find a constraint of the form ``in
  an LPB representing $\phi$ one has to ensure that $a_i \geq w \cdot a_j$''.

It has to be said however that there have been previous attempts to
somehow directly translate the occurrence patterns into numeric
coefficients or better, coefficient ratios; the threshold recognition
problem has stubbornly resisted such attempts\footnote{Personal
  communication with Yves Crama}.

However, even a \emph{rough estimate} of the coefficient ration, based
on the occurrence patterns, might be useful for reducing if not
eliminating the backtracking effort. 

One other interesting topic is a more thorough analysis of the
complexity of the combinatorial algorithm, whether it is in its
current state or after having achieved completeness. In particular, as
we have mentioned in Subsec.~\ref{only12}, analysing the effect of
exploiting the symmetries in the input DNF would be interesting.

\paragraph{Acknowledgements} 
We thank Yves Crama and Utz-Uwe Haus for very fruitful discussions about
this work.

\bibliography{bibliography}

\begin{thebibliography}{}

\bibitem[\protect\citeauthoryear{Crama and Hammer}{2011}]{CraHam07}
Crama, Y., and Hammer, P.~L.
\newblock 2011.
\newblock {\em Boolean Functions - Theory, Algorithms, and Applications},
  volume 142 of {\em Encyclopedia of mathematics and its applications}.
\newblock Cambridge University Press.

\bibitem[\protect\citeauthoryear{Dixon and Ginsberg}{2000}]{DixGin-KER00}
Dixon, H.~E., and Ginsberg, M.~L.
\newblock 2000.
\newblock Combining satisfiability techniques from {AI} and {OR}.
\newblock {\em The Knowledge Engineering Review} 15(1):31--45.

\bibitem[\protect\citeauthoryear{Fr{\"{a}}nzle and Herde}{2007}]{FraHer06}
Fr{\"{a}}nzle, M., and Herde, C.
\newblock 2007.
\newblock {HySAT}: An efficient proof engine for bounded model checking of
  hybrid systems.
\newblock {\em Formal Methods Syst. Des.} 30(3):179--198.

\bibitem[\protect\citeauthoryear{Hooker}{1992}]{Hoo92}
Hooker, J.~N.
\newblock 1992.
\newblock Generalized resolution for 0-1 linear inequalities.
\newblock {\em Ann. Math. Artif. Intell.} 6(1-3):271--286.

\bibitem[\protect\citeauthoryear{Peled and Simeone}{1985}]{PeledSimeone85}
Peled, U.~N., and Simeone, B.
\newblock 1985.
\newblock Polynomial-time algorithms for regular set-covering and threshold
  synthesis.
\newblock {\em Discret. Appl. Math.} 12(1):57--69.

\bibitem[\protect\citeauthoryear{Schilling}{2011}]{SchillingBachelor11}
Schilling, C.
\newblock 2011.
\newblock {Solving the Threshold Synthesis Problem of {B}oolean Functions by
  Translation to Linear Programming}.
\newblock Bachelor thesis, Albert-Ludwigs-Universit\"at Freiburg.

\bibitem[\protect\citeauthoryear{Smaus}{2007a}]{Sma-CPAIOR07}
Smaus, J.
\newblock 2007a.
\newblock On {B}oolean functions encodable as a single linear pseudo-{B}oolean
  constraint.
\newblock In {\em {CPAIOR}}, volume 4510 of {\em LNCS},  288--302.
\newblock Springer.

\bibitem[\protect\citeauthoryear{Smaus}{2007b}]{SmaTR230-07nociteother}
Smaus, J.
\newblock 2007b.
\newblock On {B}oolean functions encodable as a single linear pseudo-{B}oolean
  constraint.
\newblock Technical Report 230, Institut f{\"u}r Informatik, Universit{\"a}t
  Freiburg.
\newblock Long version of \cite{Sma-CPAIOR07}. Also available as TR No.~13 on
  \url{www.avacs.org}.

\bibitem[\protect\citeauthoryear{Wegener}{1987}]{Weg87-nohtml}
Wegener, I.
\newblock 1987.
\newblock {\em The complexity of {B}oolean functions}.
\newblock Wiley-Teubner.

\bibitem[\protect\citeauthoryear{Wenzelmann}{2011}]{WenzelmannBachelor11}
Wenzelmann, F.
\newblock 2011.
\newblock {Solving the Threshold Synthesis Problem of {B}oolean Functions by a
  Combinatorial Algorithm}.
\newblock Bachelor thesis, Albert-Ludwigs-Universit\"at Freiburg.

\bibitem[\protect\citeauthoryear{Winder}{1962}]{winder62}
Winder, R.~O.
\newblock 1962.
\newblock {\em Threshold Logic}.
\newblock Ph.D. Dissertation, Department of Mathematics, Princeton University,
  Princeton, U.S.A.

\end{thebibliography}

\end{document}